\documentclass[preprint,12pt]{elsarticle}

\usepackage{amssymb}
\usepackage{graphicx}
\usepackage{tabularx, multirow}
\usepackage{booktabs}
\usepackage{url}

\journal{}

\begin{document}

\begin{frontmatter}



\title{Synthesizing Bidirectional Temporal States of Knee Osteoarthritis Radiographs with Cycle-Consistent Generative Adversarial Neural Networks}

\author[1]{Fabi Prezja}
\author[3,4]{Leevi Annala}
\author[1]{Sampsa Kiiskinen}
\author[1,5]{Suvi Lahtinen}
\author[1]{Timo Ojala}

\affiliation[1]{organization={University of Jyväskylä, Faculty of Information Technology},
            city={Jyväskylä},
            postcode={40014}, 
            country={Finland}}

\affiliation[3]{organization={University of Helsinki, Faculty of Science, Department of Computer Science},
            city={Helsinki},
            country={Finland}}

\affiliation[4]{organization={University of Helsinki, Faculty of Agriculture and Forestry, Department of Food and Nutrition},
            city={Helsinki},
            country={Finland}}

\affiliation[5]{organization={University of Jyväskylä, Faculty of Mathematics and Science, Department of Biological and Environmental Science},
            city={Jyväskylä},
            postcode={40014}, 
            country={Finland}}

\begin{abstract}
Knee Osteoarthritis (KOA), a leading cause of disability worldwide, is challenging to detect early due to subtle radiographic indicators. Diverse, extensive datasets are needed but are challenging to compile because of privacy, data collection limitations, and the progressive nature of KOA. However, a model capable of projecting genuine radiographs into different OA stages could augment data pools, enhance algorithm training, and offer pre-emptive prognostic insights. In this study, we trained a CycleGAN model to synthesize past and future stages of KOA on any genuine radiograph. The model was validated using a Convolutional Neural Network that was deceived into misclassifying disease stages in transformed images, demonstrating the CycleGAN's ability to effectively transform disease characteristics forward or backward in time. The model was particularly effective in synthesizing future disease states and showed an exceptional ability to retroactively transition late-stage radiographs to earlier stages by eliminating osteophytes and expanding knee joint space, signature characteristics of None or Doubtful KOA. The model's results signify a promising potential for enhancing diagnostic models, data augmentation, and educational and prognostic usage in healthcare. Nevertheless, further refinement, validation, and a broader evaluation process encompassing both CNN-based assessments and expert medical feedback are emphasized for future research and development.
\end{abstract}


\end{frontmatter}


\newcommand{\qvec}[1]{\textbf{\textit{#1}}}

\section{Introduction}
The integration of artificial intelligence into the medical field has seen a rapid acceleration over the past decade\cite{wang2019deep,beam2018big}, due, in part, to the expansive growth of deep machine learning techniques\cite{lecun2015deep}. Notable progress in the realms of diagnostics and disease classification has been documented, as evidenced by several works\cite{esteva2017dermatologist,han2017breast,bakator2018deep,prezja2023improved,prezja2023improving}. One application of these developments is the generation and anonymization with synthetic data\cite{chuquicusma2018fool,calimeri2017biomedical,frid2018gan,thambawita2021deepfake,annala2020generating,shin2018medical,yoon2020anonymization,torfi2022differentially,kasthurirathne2021generative,prezja2022synthetic, prezja2022deepfake}, a field which has shown promise but also exposed challenges, particularly in the domain of Osteoarthritis (OA)\cite{prezja2022deepfake}.

OA, especially Knee joint osteoarthritis (KOA)\cite{HUNTER20191745,yeoh2021emergence,saarakkala2010depth, laasanen2003biomechanical}, is a leading cause of disability worldwide\cite{hermans2012productivity}, with expected costs amounting to nearly 2.5\% of the Gross National Product in western countries\cite{hermans2012productivity}. Subtle radiographic indicators and the variability in disease progression hamper the early detection of KOA\cite{HUNTER20191745,yeoh2021emergence}. In our previous study \cite{prezja2022deepfake}, synthetic radiographs were employed as a successful tool for data augmentation, creating additional diagnostic data to train deep learning models and enhancing their capacity to recognize diverse KOA presentations. This demonstrated a compelling use case for synthetic radiographs. However, it also highlighted an unmet need - the ability to modify existing KOA radiographs to represent past or future states of the disease.

The utility of deep learning in KOA diagnosis\cite{tiulpin2019multimodal,tiulpin2018automatic,tiulpin2020automatic}is highly reliant on the availability of diverse and extensive datasets. However, acquiring such datasets can be challenging due to patient privacy\cite{centers2003hipaa,voigt2017eu}, data collection constraints, and the nature of OA's progression. In this context, a tool that can alter existing genuine or synthetic radiographs into different OA stages could substantially augment the existing pool of data, making the training of diagnostic algorithms improved and more robust\cite{frid2018gan,calimeri2017biomedical,ge2019cross,mok2018learning,bowles2018gan,madani2018chest}.

Beyond data augmentation, another critical area of need is prognostic modeling. Predicting the future state of a progressive disease like OA is complex\cite{hochberg1996prognosis} but crucial for effective patient management. By using neural networks to transform current radiographs to reflect future disease states, we could potentially predict and visualize the disease trajectory, providing invaluable information for prognosis and treatment planning.

Building upon our previous work\cite{prezja2022deepfake}, our current study presents a novel approach, introducing a Cycle-Consistent Generative Adversarial Network (CycleGAN)\cite{zhu2017unpaired} capable of synthesizing the progression and regression of KOA on authentic radiographic images. This supervised, bidirectional approach is the first in the field to synthesize future and past states of KOA on existing radiographs. This study sets the stage for subsequent research into augmentation techniques and prognostic applications leveraging this advanced synthesis capability.

\section{Methods}
\begin{figure}[!ht]
\centering
\includegraphics[width=\textwidth ]{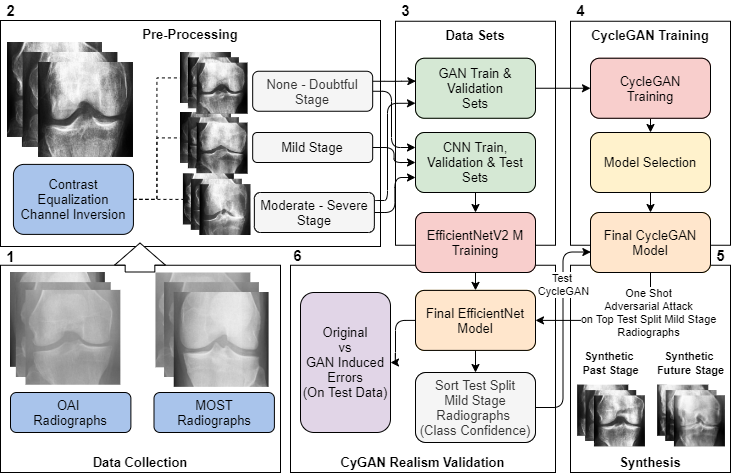}
\caption{Flowchart illustrating the tasks and data pertinent to the study. Data components are highlighted in blue and green, while model training is indicated in red. Criteria-based sorting is denoted in gray, and the final models are showcased in orange.Numeric markers indicate the order of operations.}
\label{fig:flows}
\end{figure}

Methods are categorized into three parts: The initial part focuses on data collection and pre-processing. The middle section elaborates on training generative adversarial neural networks and convolutional neural networks. The concluding segment is dedicated to the realism validation of the CyGAN. Figure \ref{fig:flows} summarizes and breaks down these components into small steps.

\begin{figure}[!ht]
\centering
\includegraphics[width=\textwidth ]{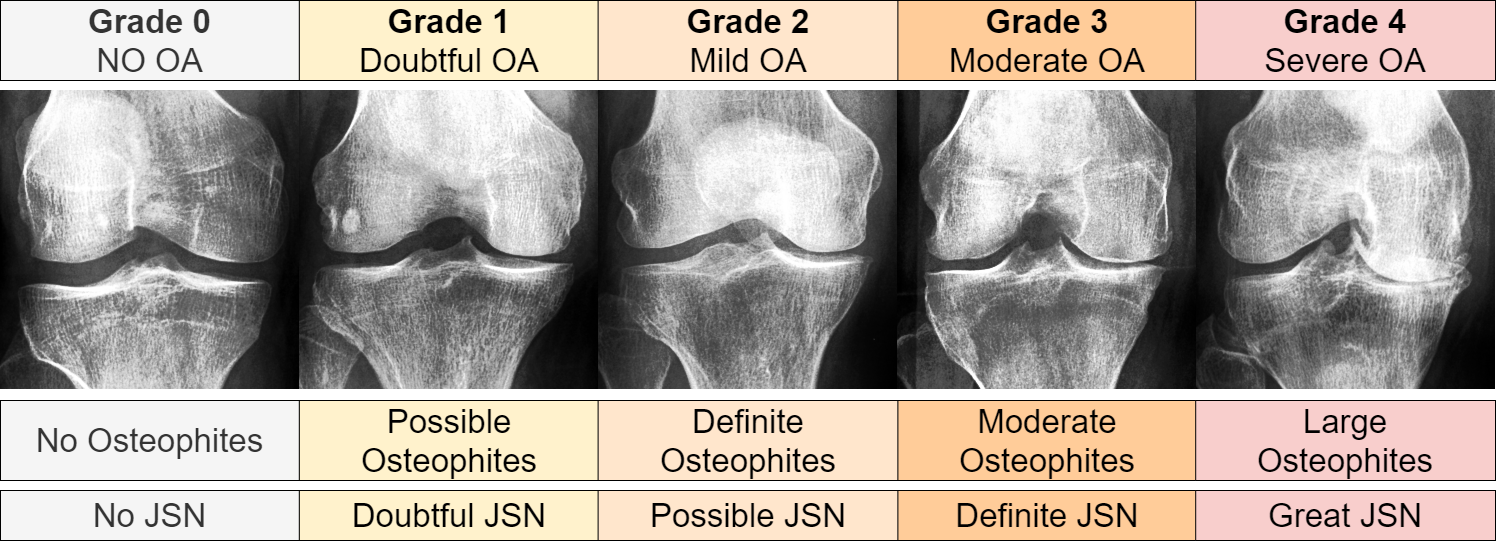}
\caption{Random examples across all KL grades, demonstrating their primary criteria. From left to right, KL grades increase from 0 (no radiological signs of OA) to 4 (severe OA). Joint space narrowing, denoted as JSN.}
\label{fig:grades}
\end{figure}

Our research began with the acquisition of knee joint radiograph images from Chen 2019\cite{chen2019fully}, leveraging data from the Osteoarthritis Initiative (OAI)\cite{nevitt2006osteoarthritis}, and the Multicenter Osteoarthritis Study (MOST) dataset\cite{segal2013multicenter,segal2013multicente1}. These separate data sets were integrated to construct an extensive dataset for our training purposes. The OAI data, originating from a multi-center longitudinal study, involved 4796 participants aged 45 to 79 years. On the other hand, the MOST dataset encompassed 3026 participants at baseline, we constructed a new dataset based on the MOST data processed by Tuomikoski\cite{tuomikoski2023unsupervised}. The new set contained adjustments for inconsistent image resolutions, contrast, antiallazing, negative radiographs, partially obstrctructed radiographs, and knee false-positive radiographs (not containing any knee-joints). For both sets the aggregated data resulted in a dataset comprising of 14134 individual knee joint images, each with a size of 224 x 224 pixel with a standardized zoom level of 0.14mm/pixel. Radiographs were graded using the Kellgren and Lawrence system\cite{kellgren1957radiological}. Figure \ref{fig:grades} illustrates an enhanced contrast example for each KL grade. For our study, the ground truth labels were categorized as 'None - Doubtful Stage' for KL0 and KL1, 'Mild Stage' for KL2, and 'Moderate - Severe Stage' for KL3 and KL4. This classification is also clinically relevant. In conclusion, the None-Doubtful stage group contained 8278 images, the Mild stage group 3100 images, and the Moderate-Severe stage group comprised the remaining 2756 images.

The CycleGAN was trained on two classes: 'None - Doubtful Stage' and 'Moderate - Severe Stage'. We reserved the 'Mild' middle stage images specifically for testing forward and backward propagation. Meanwhile, the CNN was trained using all available classes. For both the CNN and CycleGAN training, the data was partitioned as detailed in each respective section. Only previously unseen test images from both the CNN and CycleGAN were employed for result generation. We utilized the Python libraries Tensorflow, Deep Fast Vision, and Numpy for the training processes. Scikit-image, another Python library, played a pivotal role in image processing tasks.

\subsection{Image Pre-Proceessing}

After merging the KL levels, the subsequent phase involved a series of image processing steps. Initially, we performed a lateral flip on each right knee joint orientation, aligning it to the left. Subsequently, we identified and inverted negative channel images, finding 189 instances. Following the channel inversion process, we employed contrast equalization on the histograms of the images. The technique is detailed in Equation \ref{eq:contr_norm}. Given a gray-scale image $\qvec{I}$ of dimensions $m\times n $, we utilized its cumulative distribution function (cdf) and pixel value $v$ to attain an equalized value $h(v)$ within the range $[0,255]$:

\begin{equation}\label{eq:contr_norm}
h(v)=255\frac{\mathrm{cdf}(v)-\mathrm{cdf}_{\mathrm{min}}}{{(mn)-\mathrm{cdf}_{\mathrm{min}}}}
\end{equation}

Here, $\mathrm{cdf}_{\mathrm{min}}$ represents a non-zero minimum value of the image's cumulative distribution, and $mn$ indicates the total pixel count.

\subsection{Cycle-Consistent Generative Adversarial Neural Network}

We employed a Cycle-Consistent Generative Adversarial Neural Network\cite{zhu2017unpaired} (CycleGAN) for training, using the 'None - Doubtful Stage' and 'Moderate - Severe Stage' images as target domains. Images classified as 'Mild Stage' were utilized as a follow-up validation set after training. CycleGANs, are a specialized type of Generative Adversarial Network (GAN)\cite{goodfellow2014generative} designed for mapping translation between two unpaired data domains. The unique architecture of a CycleGAN includes two generator networks ($G$ and $F$), and two discriminator networks ($D_Y$ and $D_X$). The generators ($G$ and $F$) translate data between one domain and the other and vice versa. In contrast, the discriminators ($D_Y$ and $D_X$) distinguish real images from those generated in their respective domains. The CycleGAN model utilizes a combined loss function comprising both adversarial and cycle consistency losses. The adversarial loss ensures that the generated images appear realistic, while the cycle consistency loss ensures that an image translated to the other domain and subsequently reversed yields the original image. This is represented mathematically as follows:

\begin{equation}
L(G, F, D_X, D_Y) =  L_{GAN}(G, D_Y, X, Y) + L_{GAN}(F, D_X, Y, X) 
 + \lambda L_{cyc}(G, F)
\end{equation}

\begin{figure}[!ht]
\centering
\includegraphics[width=\textwidth ]{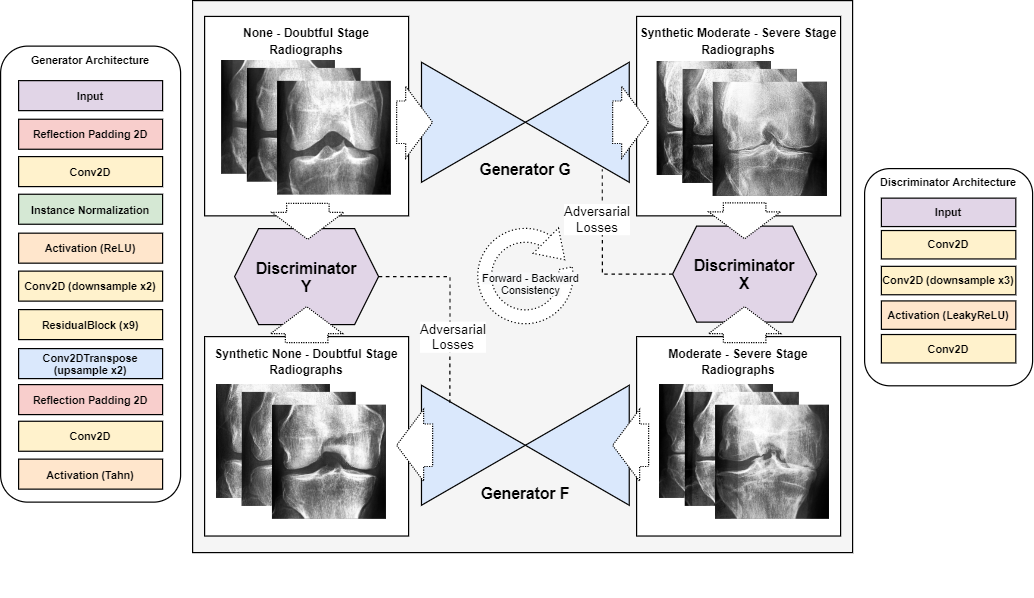}
\caption{In-depth depiction of the training process and structural aspects of our CycleGAN Model. All generators and discriminators uniformly adhere to the portrayed architectures.}
\label{fig:flowgan}
\end{figure}

For the generators $G$ and $F$, $L_{GAN}(G, D_Y, X, Y)$ and $L_{GAN}(F, D_X, Y, X)$, provide the adversarial losses. The cycle consistency loss, $L_{cyc}(G, F)$, is a separate term in the loss function. Here, $\lambda$ represents a hyperparameter that controls the weight of the cycle consistency loss. The cycle consistency loss can be further detailed as:

\begin{equation}
L_{cyc}(G, F) =  \mathbb{E}_{x\sim p_{data}(x)}[||F(G(x))-x||_1] 
 + \mathbb{E}_{y\sim p_{data}(y)}[||G(F(y))-y||_1]
\end{equation}

where the first term represents the consistency loss for domain $X$ and the second term for domain $Y$. CycleGANs facilitate unpaired image-to-image translation tasks, which are practical for various applications in computer vision. A graphical representation of our basic architectural blocks and processes can be seen in Figure \ref{fig:flowgan}. Additionally, the specifics of our CycleGAN architecture are detailed in tables \ref{tab:generatorarch}, \ref{tab:discriminatorarch}, and \ref{table:blocksarch}. Each generator in our model was composed of 11,370,881 parameters, while the discriminators were slightly leaner, comprising 11,032,193 parameters each.

\subsubsection{Experiment Architecture and parameters.}

The CycleGAN model was trained on 90\% of the data and 10\% for validation (patient aware) using the Adam optimizer (learning rate 2e-4, betas 0.5 and 0.999) on both generators and discriminators. We used a batch size of 8 and underwent training for 155 epochs. The CycleGAN architecture consisted of two generators and two discriminators, with the generators employing downsampling, residual\cite{he2016deep,targ2016resnet}, and upsampling blocks and the discriminators utilizing downsampling blocks. The blocks are further detailed in Table \ref{table:blocksarch}. The model's loss functions included adversarial loss, cycle consistency loss, and identity loss, with the latter two governed by lambda values of 10 and 0.5, respectively. The duration of training was roughly two days with a P100 GPU. The model with the lowest loss for both generators was chosen as the final model (epoch 101). Saliency maps were generated by overlaying the GAN-generated image (using the the 'inferno' Matplotlib\cite{hunter2007matplotlib} colormap) onto the original image (using the reversed colormap) with a 0.4 transparency.

\subsection{CyGAN Validation}

To validate our model, we employed a Convolutional Neural Network (CNN)\cite{lecun1995convolutional} designed to differentiate between all classes of the Knee Osteoarthritis (KOA) problem (detailed bellow). Our ultimate aim was to fool this CNN into categorizing images into a class of our choice. We applied this strategy by using unseen 'Mild Stage' test images, which were not used in either the GAN or the CNN training. Firstly, we used the CNN to rank 100 of these test images according to their class confidence level for belonging to the 'Mild Stage.' Next, we fed these top 100 images to both generators within our CycleGAN. The generators produced two new sets of images, reflecting transformations to both ends of the KOA spectrum: the 'None - Doubtful Stage' and the 'Moderate - Severe Stage.' To assess the efficacy of our CycleGAN, we used the CNN to predict the classes of these transformed images and subsequently analyzed the results. We repeated this procedure for GAN and CNN shared test images from the 'None - Doubtful Stage' and 'Moderate - Severe Stage' classes. We considered this a 'one-shot' adversarial attack because we aimed to drastically alter the CNN's classification with a single application of our CycleGAN.

\subsection{EfficientNetV2 Convolutional Neural Network}

\begin{figure} [!ht]
\centering
\includegraphics[width=\textwidth ]{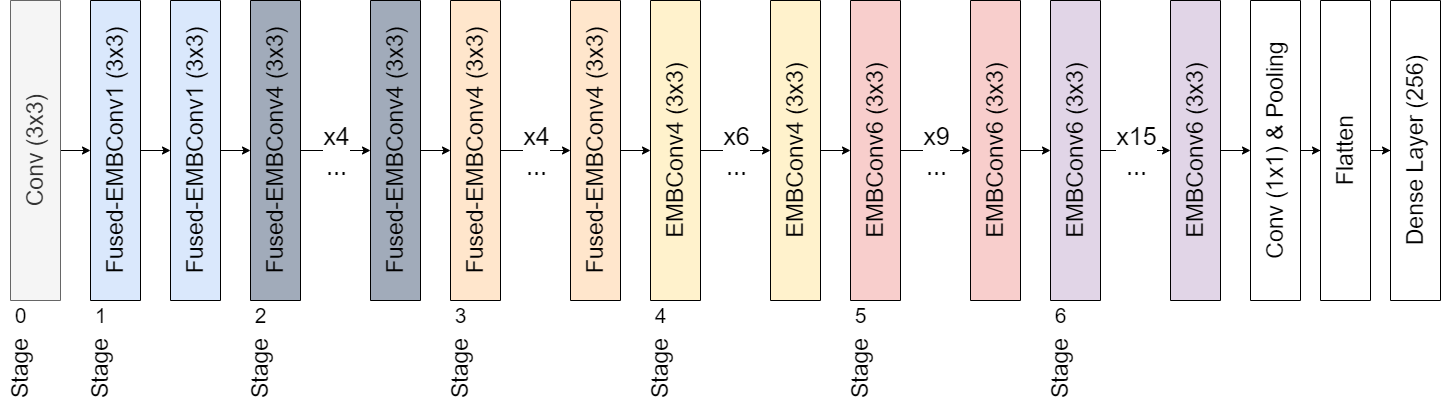}
\caption{Integrated architecture of the base EfficientNetV2, further detailing our specific modifications implemented post the pooling layer.} 
\label{fig:effinet}
\end{figure}

EfficientNet\cite{tan2019efficientnet} is a novel model in the domain of deep learning that has earned significant acclaim for its effectiveness. The EfficientNet framework is engineered to uniformly scale all dimensions, including depth, width, and resolution. At the heart of EfficientNet lies the principle of compound scaling. This approach strives to strike a balance between the network's depth (number of layers), width (the size of the layers), and resolution (size of the input image). A set of fixed scaling coefficients governs this balance. Specifically, for a baseline EfficientNet-B0 (\ref{fig:effinet}), if $\alpha, \beta, \gamma$ are constants to maintain the equilibrium, and $\phi$ is the coefficient defined by the user, the depth $d$, width $w$, and resolution $r$ of the network can be scaled as follows:

\begin{equation}\label{eq:enet}
d = \alpha^{\phi} d_0, \quad w = \beta^{\phi} w_0, \quad r = \gamma^{\phi} r_0
\end{equation}
In this equation, $d_0, w_0, r_0$ signify the base model's depth, width, and resolution, respectively.

An integral part of EfficientNet is the MBConv block, which takes inspiration from the MobileNetV2 architecture\cite{sandler2018mobilenetv2}. This block initiates a sequence of transformations: a $1\times1$ convolution (expansion), followed by a depth-wise convolution (indicated by a depth-wise separable convolution kernel $\qvec{D}$), a Squeeze-and-Excitation (SE) operation\cite{hu2018squeeze} , and another $1\times1$ convolution (projection). For an input image $\qvec{I}$, the transformation of the MBConv block, $T_{MB}$, can be depicted as:

\begin{equation}\label{eq:mbconv_se}
T_{MB}(\qvec{I}) = \qvec{K}_2 \ast SE(\qvec{D} \ast (\qvec{K}_1 \ast \qvec{I}))
\end{equation}

In this formula $\ast$ denotes the convolution operation, $\qvec{K}_1$ and $\qvec{K}_2$ are the $1\times1$ convolutional filters, $\qvec{D}$ illustrates the depth-wise convolutional filter, and $SE(\cdot)$ stands for the Squeeze-and-Excitation operation. An activation function succeeds each convolution. The efficient employment of computational resources within the MBConv block, particularly via the depth-wise convolution and the SE block, significantly contributes to EfficientNet's superior performance.

EfficientNetV2\cite{tan2021efficientnetv2} brings several pivotal changes to the original EfficientNet blueprint. The depth, width, and resolution scaling remain identical to the original EfficientNet. However, the introduction of the Fused-MBConv block, which merges the initial $1\times1$ convolution and the depth-wise convolution into a single $13\times3$ convolution operation, followed by the Squeeze-and-Excitation (SE) operation and a final $1\times1$ convolution, marks a significant adjustment.

The transformation of the Fused-MBConv block, $T_{FMB}$, can be represented as:

\begin{equation}\label{eq:fmbconv}
T_{FMB}(\qvec{I}) = \qvec{K}_{2} \ast (SE(\qvec{K}_{f} \ast \qvec{I}))
\end{equation}

In this formula, $\qvec{K}_{f}$ is the $3\times3$ convolutional filter that combines the initial $1\times1$ convolution and the depth-wise convolution, $\qvec{K}_{2}$ is the concluding $1\times1$ convolutional filter, and $SE(\cdot)$ denotes the Squeeze-and-Excitation operation. Each convolution, followed by an activation, may be in a block with a skip connection.

In our study, we utilized the EfficientNetV2-M model and augmented it by adding a flattening layer followed by a dense layer with 256 neurons. This dense layer employed an Exponential Linear Unit\cite{clevert2015fast} (ELU) as its activation function, placed before the output layer. During the training phase, we used Categorical Cross-Entropy Loss and conducted training over 15 epochs. We divided the dataset into Training, Validation, and Testing sets, allocating 70\%, 15\%, and 15\% respectively (patient aware). This CNN was trained across all three KOA stages. Early stopping was dictated by the minimum validation loss with a tolerance set at 4. The entire model training procedure was carried out using the Deep Fast Vision repository\cite{fabprezja_2023}. We used the Adam\cite{kingma2014adam} optimizer for the task, with a learning rate of $2 \times 10^{-5}$ and a batch size of 32. Furthermore, we enabled the advanced automatic augmentation capabilities from the library.

\subsection{CNN Interpretability}
While the complexity of neural networks increases their capabilities, it also complicates the interpretation of their predictions\cite{prezja2023importance}. Due to this complexity, these systems are often deemed as 'black boxes.' However, the Grad-CAM\cite{selvaraju2017grad} algorithm (based on the CAM\cite{Oquab_2015_CVPR} framework) helps demystify this 'black box' effect. At a high level, Grad-CAM is an algorithm that visualizes how a convolutional neural network makes its decisions. It creates what are known as "heat maps" or "activation maps" that highlight the areas in an input image that the model considers important for making its prediction.

\section{Results}

\subsection{CyGAN Training Results}
In this study, we developed a CycleGAN model that can synthesize osteoarthritis's progression or regression on genuine radiographs. We successfully validated our model by using a Convolutional Neural Network (CNN) called EfficientNetV2M. Our primary objective was to effectively transform 'Mild Stage' osteoarthritis images using our CycleGAN model, causing the CNN to misclassify the stage of the disease on demand. This confirmed our CycleGAN model's effectiveness in transforming images to represent either an earlier or later stage of the disease.

We divided our results into three sections for clarity. The first section demonstrates all model training results. The second section highlights top examples from the test set that were used to fool the CNN's osteoarthritis model. Finally, we globally visualize test images via the t-distributed stochastic neighbor embedding method\cite{van2008visualizing}.

\begin{figure} [!ht]
\centering
\includegraphics[width=\textwidth]{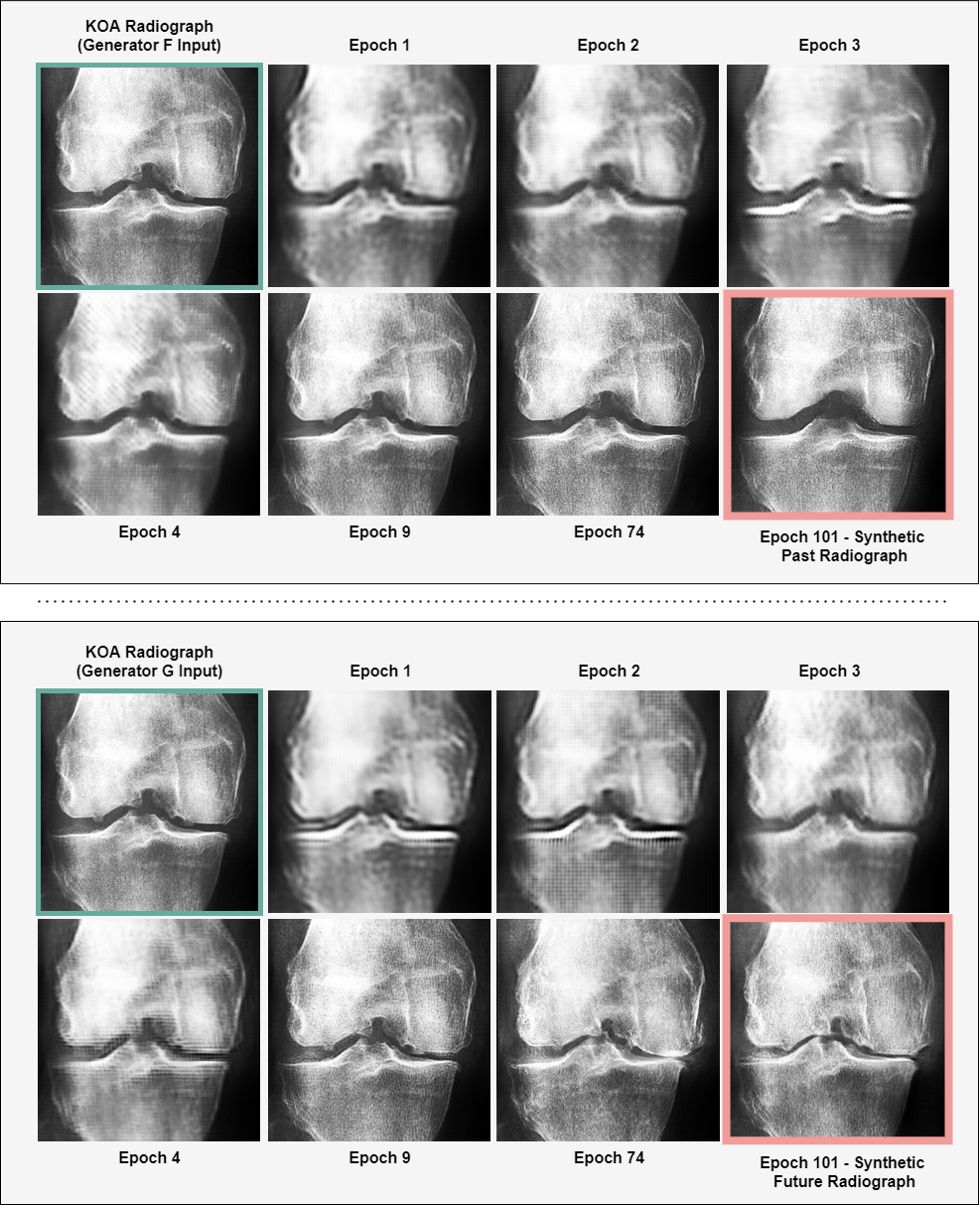}
\caption{Results from querying the stored models using a single test image up to the optimal epoch. It includes outputs for both CycleGAN generators.} 
\label{fig:trains}
\end{figure}

As illustrated in Figure \ref{fig:trains}, there was a substantial enhancement in the quality of the synthetically generated states as the training of the CycleGAN progressed. Notably, once the generators mastered the initial representation of the input images, we observed the emergence of more significant transformations. This suggests an effective learning result within the CycleGAN, as it appears to have first developed an understanding of the input before executing more complex and impactful modifications.

\subsection{CNN Training Results}
\begin{figure} [!ht]
\includegraphics[width=\textwidth]{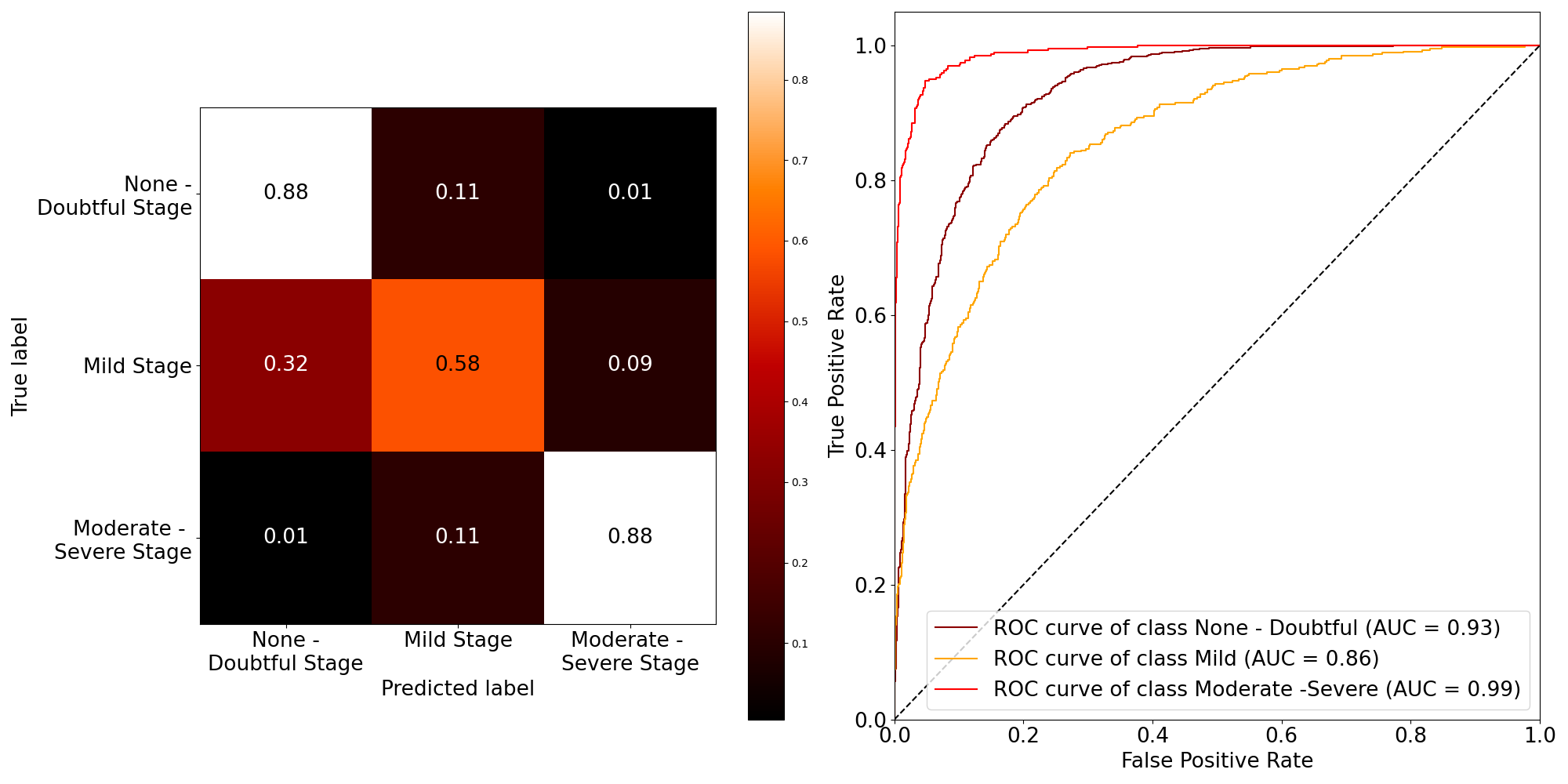}
\caption{Confusion matrix and ROC curves for the CNN classifier. ROC curves are constructed using a one-vs-all approach.}
\label{fig:rocs}
\end{figure}

Before we delve into the outcomes produced by our CycleGAN and its capacity to manipulate osteoarthritic radiographs, it is crucial to firstunderstand the performance of our base model, the Convolutional Neural Network (CNN) for Knee Osteoarthritis (KOA) classification. This is the model we attempted to 'fool' through synthetic images produced by the CycleGAN. Figure \ref{fig:rocs} presents the confusion matrix and AUCs of the model, while table \ref{tab:classification_report} presents some key metrics of its performance.

\begin{table}[!ht]
\caption{Classification report, detailing precision, recall, and F1-score for each class. It also provides weighted and macro averages. The overall average accuracy of the classifier was 0.817.}
\label{tab:classification_report}
\small
\begin{tabularx}{\textwidth}{lXXXXX}
\toprule
\textbf{Metrics} & \textbf{None-Doubtful Stage} & \textbf{Mild Stage} & \textbf{Moderate-Severe Stage} & \textbf{Weighted Average Metric} & \textbf{Macro Average Metric} \\
\midrule
Precision & 0.877 & 0.601 & 0.873 & 0.815 & 0.784 \\
Recall    & 0.882 & 0.584 & 0.884 & 0.817 & 0.784 \\
F1-score  & 0.880 & 0.593 & 0.879 & 0.816 & 0.784 \\
\bottomrule
\end{tabularx}
\end{table}

The classification report table \ref{tab:classification_report} encapsulates our model's effectiveness in identifying different stages of osteoarthritis. High precision and recall values for 'None - Doubtful Stage' and 'Moderate - Severe Stage' (approximately 0.88) indicate accurate predictions and minimal false negatives for these categories. 

\subsection{Adversarial Attacks \& Synthetic States}

Table \ref{tab:adversarl} in the study represents an empirical demonstration of the effectiveness of our method in misleading the CNN classifier, EfficientNetV2. This was achieved by applying the adversarial attack through our CycleGAN model, which altered the test images to misrepresent the original state of knee osteoarthritis.The table displays the original class of the radiographs (None-Early Stage, Moderate - Severe Stage, and Mild Stage), the direction of the adversarial attack (either towards the future, simulating progression, or towards the past, simulating regression of the disease), and the consequent CNN prediction across the three categories (None-Early Stage, Mild Stage, and Moderate - Severe Stage).

\begin{table}[!ht]
\caption{The impact of an adversarial one-shot attack on the trained classifier using test images. It specifically showcases the shifts in classification outcomes resulting from the application of CycleGAN.}
\label{tab:adversarl}
\tiny
\begin{tabularx}{\textwidth}{lXXXX}
\toprule
\textbf{Original Class} & \textbf{Adversarial Attack Direction} & \textbf{CNN Prediction (None - Doubtful Stage)} & \textbf{CNN Prediction (Mild Stage)} & \textbf{CNN Prediction (Moderate - Severe Stage)} \\
\midrule
None - Doubtful Stage & Towards Future & 5.13\% & 11.11\% & 83.76\% \\
Moderate - Severe Stage & Towards Past & 75.61\% & 19.51\% & 4.88\% \\
Mild Stage & Towards Future & - & 24.00\% & 76.00\% \\
Mild Stage & Towards Past & 31.00\% & 69.00\% & - \\
\bottomrule
\end{tabularx}
\end{table}

For the 'None - Doubtful Stage' radiographs, when the adversarial attack was directed towards the future, the CNN prediction was largely shifted to the 'Moderate - Severe Stage' (83.76\%). Only a minor fraction was identified as 'None - Doubtful Stage' (5.13\%) or 'Mild Stage' (11.11\%), indicating a substantial alteration in classification due to the attack. The 'Moderate - Severe Stage' radiographs displayed a similar trend. When manipulated towards the past, a considerable percentage (75.61\%) was misclassified as 'None - Doubtful Stage', diverging significantly from the original classification. When the adversarial attack was applied to the 'Mild Stage' images towards the future, a majority (76.00\%) was misclassified as 'Moderate - Severe Stage'. Conversely, when the attack was directed towards the past, the 'Mild Stage' images were partly misclassified as 'None - Doubtful Stage' (31.00\%).

\begin{figure} [!ht]
\centering
\includegraphics[width=\textwidth]{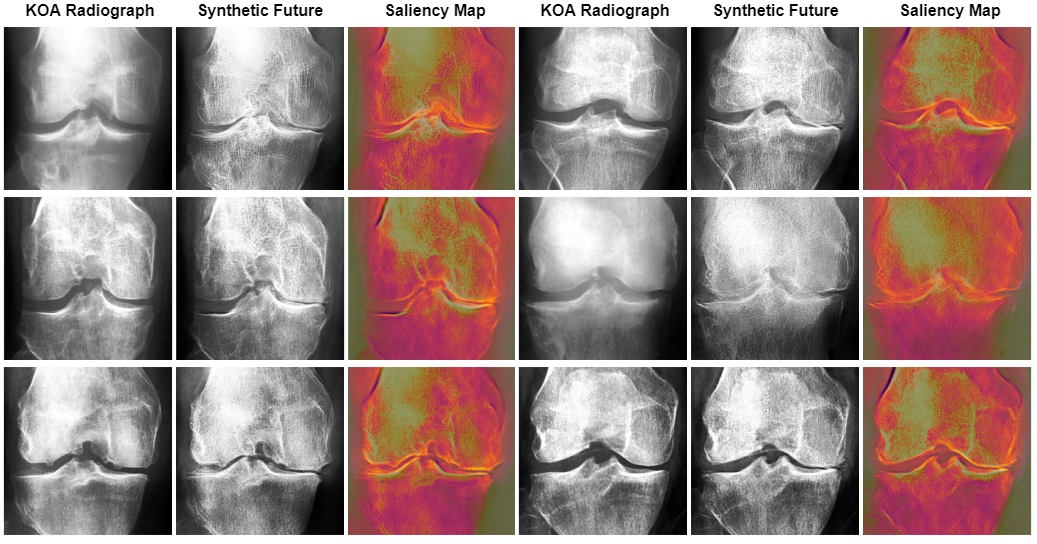}
\caption{Top 6 adversarial examples for synthetic futures with a minimum confidence level above 0.9 in the desired class. The 'KOA Radiograph' column displays the original test image selected for its high confidence in its initial class. The saliency map underscores the alterations made by our CycleGAN, with bright orange indicating additive changes and deep purple suggesting the reverse.} 
\label{fig:futures}
\end{figure}

\begin{figure} [!ht]
\centering
\includegraphics[width=\textwidth]{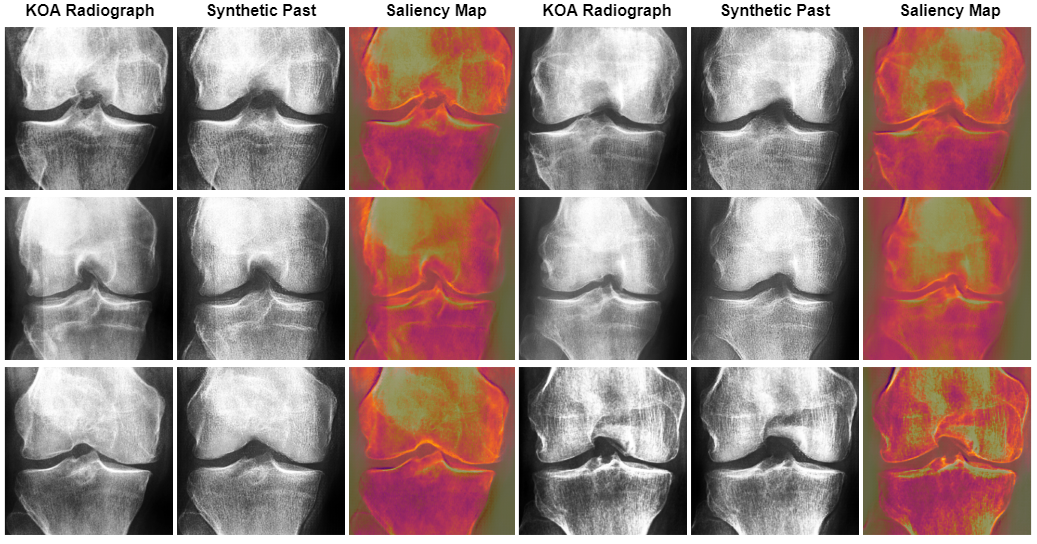}
\caption{Top 6 adversarial examples for synthetic pasts with a minimum confidence level above 0.9 in the desired class. The 'KOA Radiograph' column displays the original test image selected for its high confidence in its initial class. The saliency map underscores the alterations made by our CycleGAN, with bright orange indicating pixel diminishing changes and deep purple suggesting the reverse.} \label{fig:pasts}
\end{figure}

Figures \ref{fig:futures} and \ref{fig:pasts} present the top six test images transformed by our CycleGAN. Initially, the CNN classifier was correct and over 90\% confident about the osteoarthritis stage represented in these images (Mild stage). However, the classifier was tricked into assigning them an altered stage of the disease, once again with more than 90\% confidence in the altered stage.

Figure \ref{fig:futures} synthetic images presented the occurrence of joint space narrowing, a key indicator of progressing osteoarthritis. Notably, the CycleGAN model also emphasized or created osteophytes. These bony growths are characteristic of advanced osteoarthritis.

In contrast to Figure \ref{fig:futures}, figure \ref{fig:pasts} presents a marked expansion of the knee joint, an attribute commonly associated with no or early-stage osteoarthritic knees. In addition, the possible osteophytes - bony outgrowths typically found in osteoarthritic joints - seemed to be systematically removed by the CycleGAN model. The presence of osteophytes and joint space reduction are hallmark indications of advancing osteoarthritis. Thus, their removal and the expansion of the joint space suggested a transition towards an earlier state of the disease.

\begin{figure} [!ht]
\centering
\includegraphics[width=\textwidth]{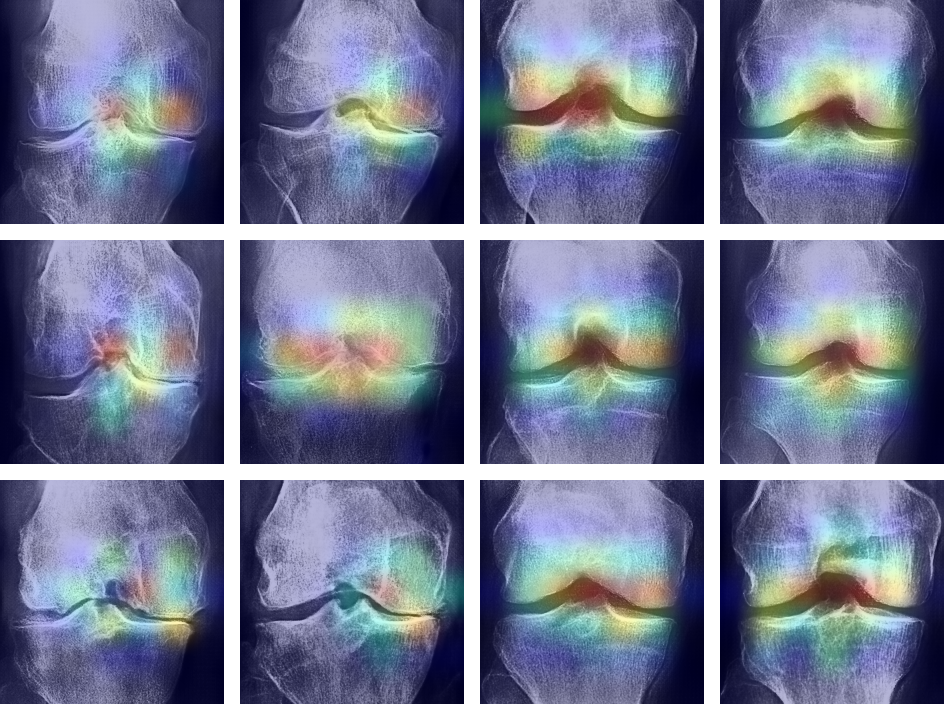}
\caption{Grad-CAM activations at the pre-flattening layer of the CNN, illustrating the top 12 adversarial images. Grad-CAM activations corresponding to the desired class are shown.} 
\label{fig:grads}
\end{figure}

Upon examining Figure \ref{fig:grads}, it was evident that the Convolutional Neural Network (CNN) model focused on key areas in the images that were modified by the Cycle Consistent Generative Adversarial Network (CyGAN). This demonstrated that the CNN based its decisions on the essential features that the CyGAN altered. This showed not only the effectiveness of the CyGAN in generating meaningful features, but also highlighted the CNN's focus on the relevant regions of interest.

\subsection{Synthetic States Visualization}

\begin{figure} [!ht]
\centering
\includegraphics[width=\textwidth]{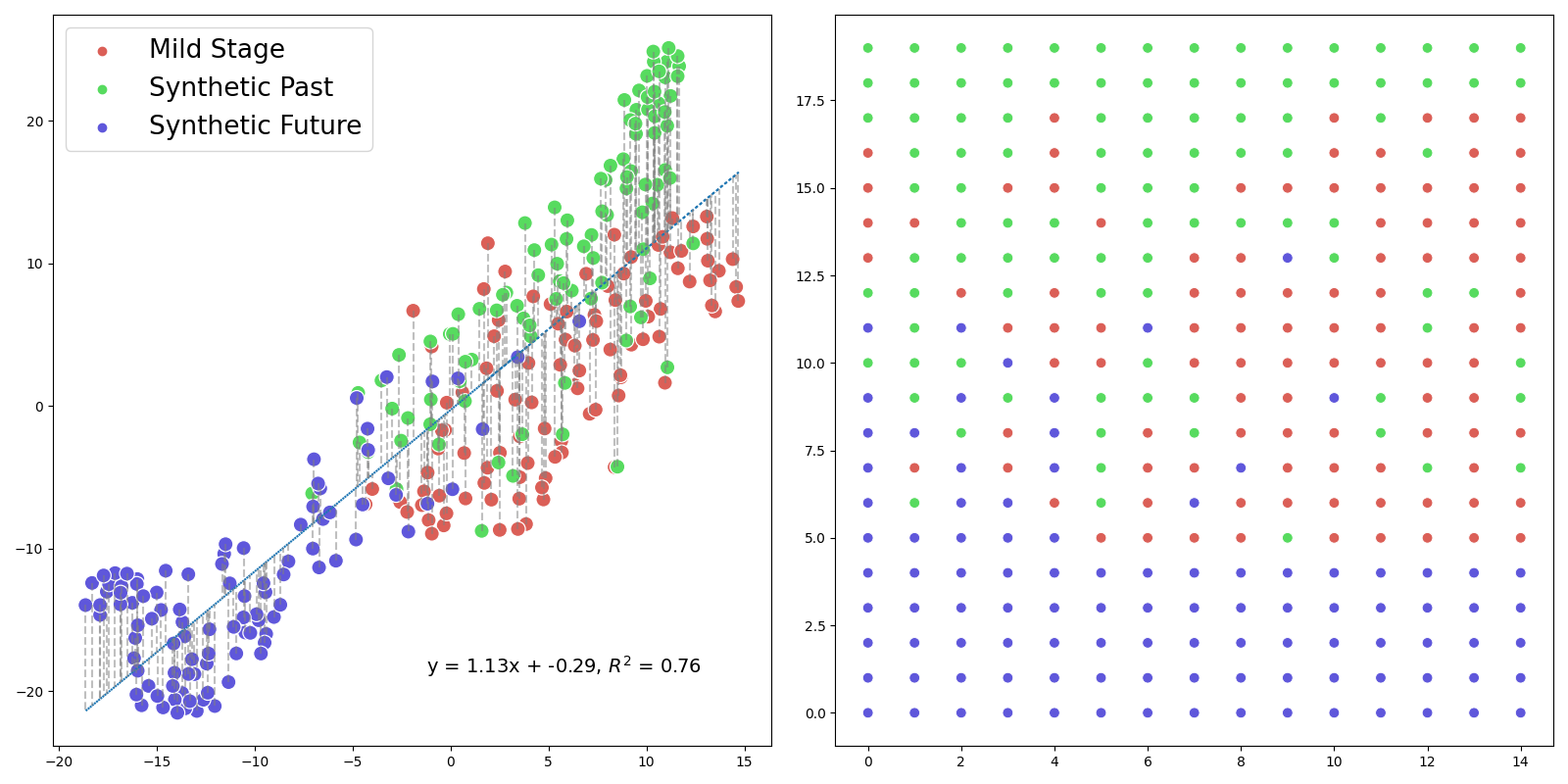}
\caption{t-SNE and Rasterized TSNE visualizations of all test and original images. A simple linear model has been fitted to the entire t-SNE cohort for ease of analysis.} 
\label{fig:tsne}
\end{figure}

By leveraging the features identified by the CNN, we've employed t-SNE\cite{van2008visualizing} to craft a visualization that projected both the original and transformed test images onto a shared space. Our aim with this method was to discern possible overlaps between synthetic and authentic radiographs.This mapping, depicted in Figure \ref{fig:tsne}, provided a compelling visualization of the connections between different image states. A rasterized rendition\cite{kogan2016ofxtsne,klingemann2016rasterfairy} of this t-SNE plot further refined our understanding, offering an enhanced view of these intricate relationships. Looking at the figure, a notable observation was the appearance of a pseudo-linear diagonal, marking an intriguing transition from early to middle, and eventually to late stages of the image states. This observation was particularly striking as it suggested a structured, semi-ordered transition. As anticipated, overlap was observed in the representation, indicating shared features or similarities among different image states. This overlap was not surprising and reaffirmed the inherent interconnectedness of these transformations.

\section{Discussion}

In our study, we created a Cycle-Consistent Generative Neural Network, designed to generate synthetic projections of genuine osteoarthritis radiographs into past and future disease states while maintaining key anatomical constraints of each radiograph. To assess its efficacy, we performed adversarial attacks on the EfficientNetV2-M architecture,  trained on a comprehensive knee osteoarthritis dataset. Our results revealed our model's capacity to convincingly shift authentic radiographs along the disease timeline, both backward and forward. The creation of this system sets the groundwork for further research, highlighting its potential for broadening classification augmentation and enhancing prognostic capabilities.

Regarding synthetic futures, as shown in our results, the model was the most effective in simulating future states. However, due to the unpaired training images, it's challenging to validate the anatomical precision of these projections without corresponding follow-up data. Despite this, our approach showed great potential. This tool could become invaluable to healthcare professionals if forecasting accuracy can be optimized and further validated. It could augment datasets for classification - a crucial advantage considering the scarcity of late-stage radiographs - and serve educational purposes. It could assist new medical experts in comprehending and simulating the effects of osteoarthritis, further contributing to their understanding of the disease.

Regarding synthetic pasts, our model showed an impressive ability to retroactively transition late-stage radiographs to earlier stages, although it was somewhat less successful with mid-stage radiographs. This capability could significantly enhance data augmentation for classification purposes. Moreover, with high accuracy in future follow-up validation, the model might contribute to future surgical planning, such as knee replacement procedures. However, it is essential to note that while this ability showed promise, it needs further development to match the model's proficiency in simulating future disease progression.

Regarding the limitations of CyGANs, as reflected in our results, a few images did not transition as anticipated, either remaining in the same class or not progressing as far as desired. This is inherent to the CycleGAN's training process\cite{zhu2017unpaired}: the generator learns to transform source images to mimic the target, while striving to keep the transformation reversible (cycle consistency). This means that, if an image is similar to the target domain, the generator minimizes changes to maintain reversibility, sometimes resulting in minimal or near-identity transformations. This tendency emerges from the model's objective of deceiving the discriminator while preserving cycle consistency. This phenomenon was evident in some of our images, where we saw significant texture changes and few anatomical alterations. To address this, a straightforward approach might involve using an Inception\cite{szegedy2016rethinking} model to compare pre and post-transform images. By using an Euclidean metric, we could detect and quantify significant anatomical changes.

In reviewing previous approaches, we note a singular\cite{han2021image} instance of an approximate system. This model, which trained without labels and for a unidirectional transform, stands out for its innovative and commendable approach, particularly in its aim to generate synthetic follow-up radiographs for individual patients. However, the model's unidirectional focus prevented label-based transformations in both directions, thereby limiting its potential as a tool for augmenting Knee Osteoarthritis (KOA) datasets and executing targeted transformations towards specific KOA KL-grades.

During our experiment, we targeted three primary stages of the disease based on their temporal occurrence: none-doubtfu, mild, and moderate-severe stages. However, there are potential benefits to be found in a more nuanced approach. In future work, it might be beneficial to consider a model capable of handling individual subclasses of the disease. This could afford a more detailed, fine-grained method to simulate the progression and regression of osteoarthritis. However, this may prove challenging due to the intrinsic scarcity of data for some individual stages\cite{prezja2022deepfake}. Drawing on previous studies that utilized GANs for augmentation\cite{prezja2022deepfake}, this fine-grained approach could lead to a wider variety of synthetic images.

In our study, we noticed that the image data we sourced from the Osteoarthritis Initiative (OAI) significantly surpassed the clarity of those from the Multicenter Osteoarthritis Study (MOST). This was partly evident in the increased requirement for anti-aliasing measures in the MOST images. However, a potential limitation was the use of images with a size of 224x224 pixels. Although this is a common standard for neural network training, it might not be optimal in our context, as the subtle changes characterizing pre-Mild osteoarthritis might not be easily discernible at this resolution. This could lead to a less precise understanding and representation of the disease's evolution in its earlier stages. In future work, We recommend potentially larger receptive fields and resolution, which might allow an increased capacity to pick up nuanced changes in the images.

Regarding validating our CycleGAN model, we successfully applied a 'one-shot' adversarial attack to a specialized Convolutional Neural Network (CNN). This method proved effective in quantifying the ability of our CycleGAN to modify class labels associated with radiographic stages of osteoarthritis. However, it's essential to recognize that this validation technique doesn't substitute for the insight and expertise of medical professionals. Therefore, it's essential that its predictions align not only with CNN-based assessments, but also with the judgments of medical experts who understand the broader context of patient health and disease manifestation\cite{prezja2022deepfake}. In future iterations, we intend to include a more comprehensive validation process. This will involve acquiring feedback and validation from medical professionals, alongside our existing CNN-based evaluations. This integrated validation approach could help develop a more resilient and clinically applicable tool, enhancing osteoarthritis diagnosis, prognosis, and management.

\section{Conclusions}

Our research successfully leveraged a Cycle-Consistent Generative Neural Network to synthesize both past and future KOA disease states in genuine radiographs. The model showcased significant potential in simulating future disease states, though it faced challenges with unpaired training images. When attempting to revert to earlier disease phases, the model demonstrated promise, particularly with advanced OA radiographs. To validate our CycleGAN model, we utilized a 'one-shot' adversarial attack on a CNN, effectively assessing its ability to modify osteoarthritis radiograph labels. While this method is insightful, it cannot replace medical expertise. Moving forward, we aim to combine this approach with feedback from medical experts.

\section*{Data Availability}
The trained models and synthetic images from the current study are available in the Google drive repository.

The Deep Fast Vision library is available at: \url{https://github.com/fabprezja/deep-fast-vision}


 \bibliographystyle{elsarticle-num} 
 \bibliography{main}

\section*{Acknowledgements}
The authors extend their sincere gratitude to Kimmo Riihiaho, Rodion Enkel and Leevi Lind.

\section*{Author contributions statement}
Conceptualization: F. P.; 
Methodology: F. P.; 
Investigation: F. P.; 
Data curation: All authors; 
Formal analysis: All authors; 
Writing – original draft: F. P.; 
Writing – review \& editing: All authors.

\section*{Additional information}
 \textbf{Competing interests}
 All authors declare that they have no conflicts of interest.

\section{Appendix}
\begin{table}[!ht]
\centering
\caption{Generator architecture of our model, detailing each layer's type, output shape, number of parameters, filters, kernel size, strides, padding, and activation function.}
\label{tab:generatorarch}
\fontsize{5pt}{4pt}\selectfont
\begin{tabularx}{\textwidth}{lXXXXXXXX}
\toprule
\textbf{Layer Type} & \textbf{Output Shape} & \textbf{Parameter Count} & \textbf{Filters} & \textbf{Kernel Size} & \textbf{Strides} & \textbf{Padding} & \textbf{Activation} \\
\midrule
InputLayer & (224, 224, 1) & 0 & - & - & - & - & - \\
ReflectionPadding2D & (230, 230, 1) & 0 & - & - & - & - & - \\
Conv2D & (224, 224, 64) & 3136 & 64 & (7, 7) & (1, 1) & valid & Linear \\
InstanceNormalization & (224, 224, 64) & 128 & - & - & - & - & - \\
Activation & (224, 224, 64) & 0 & - & - & - & - & ReLU \\
Conv2D & (112, 112, 128) & 73728 & 128 & (3, 3) & (2, 2) & same & Linear \\
InstanceNormalization & (112, 112, 128) & 256 & - & - & - & - & - \\
Activation & (112, 112, 128) & 0 & - & - & - & - & ReLU \\
Conv2D & (56, 56, 256) & 294912 & 256 & (3, 3) & (2, 2) & same & Linear \\
InstanceNormalization & (56, 56, 256) & 512 & - & - & - & - & - \\
Activation & (56, 56, 256) & 0 & - & - & - & - & ReLU \\
ReflectionPadding2D & (58, 58, 256) & 0 & - & - & - & - & - \\
Conv2D & (56, 56, 256) & 589824 & 256 & (3, 3) & (1, 1) & valid & Linear \\
InstanceNormalization & (56, 56, 256) & 512 & - & - & - & - & - \\
Activation & (56, 56, 256) & 0 & - & - & - & - & ReLU \\
ReflectionPadding2D & (58, 58, 256) & 0 & - & - & - & - & - \\
Conv2D & (56, 56, 256) & 589824 & 256 & (3, 3) & (1, 1) & valid & Linear \\
InstanceNormalization & (56, 56, 256) & 512 & - & - & - & - & - \\
Add (Skip Connection) & (56, 56, 256) & 0 & - & - & - & - & - \\
ReflectionPadding2D & (58, 58, 256) & 0 & - & - & - & - & - \\
Conv2D & (56, 56, 256) & 589824 & 256 & (3, 3) & (1, 1) & valid & Linear \\
InstanceNormalization & (56, 56, 256) & 512 & - & - & - & - & - \\
Activation & (56, 56, 256) & 0 & - & - & - & - & ReLU \\
ReflectionPadding2D & (58, 58, 256) & 0 & - & - & - & - & - \\
Conv2D & (56, 56, 256) & 589824 & 256 & (3, 3) & (1, 1) & valid & Linear \\
InstanceNormalization & (56, 56, 256) & 512 & - & - & - & - & - \\
Add (Skip Connection) & (56, 56, 256) & 0 & - & - & - & - & - \\
ReflectionPadding2D & (58, 58, 256) & 0 & - & - & - & - & - \\
Conv2D & (56, 56, 256) & 589824 & 256 & (3, 3) & (1, 1) & valid & Linear \\
InstanceNormalization & (56, 56, 256) & 512 & - & - & - & - & - \\
Activation & (56, 56, 256) & 0 & - & - & - & - & ReLU \\
ReflectionPadding2D & (58, 58, 256) & 0 & - & - & - & - & - \\
Conv2D & (56, 56, 256) & 589824 & 256 & (3, 3) & (1, 1) & valid & Linear \\
InstanceNormalization & (56, 56, 256) & 512 & - & - & - & - & - \\
Add (Skip Connection) & (56, 56, 256) & 0 & - & - & - & - & - \\
Conv2DTranspose & (112, 112, 128) & 294912 & 128 & (3, 3) & (2, 2) & same & Linear \\
InstanceNormalization & (112, 112, 128) & 256 & - & - & - & - & - \\
Activation & (112, 112, 128) & 0 & - & - & - & - & ReLU \\
Conv2DTranspose & (224, 224, 64) & 73728 & 64 & (3, 3) & (2, 2) & same & Linear \\
InstanceNormalization & (224, 224, 64) & 128 & - & - & - & - & - \\
Activation & (224, 224, 64) & 0 & - & - & - & - & ReLU \\
ReflectionPadding2D & (230, 230, 64) & 0 & - & - & - & - & - \\
Conv2D & (224, 224, 1) & 3137 & 1 & (7, 7) & (1, 1) & valid & Tanh \\
\bottomrule
\end{tabularx}
\end{table}

\begin{table}[!ht]
\caption{Discriminator architecture of our model, detailing each layer's type, output shape, number of parameters, kernel size, strides, padding, and activation function.}
\label{tab:discriminatorarch}
\scriptsize
\begin{tabularx}{\textwidth}{lXXXXXXX}
\toprule
\textbf{Layer Type} & \textbf{Output Shape} & \textbf{Parameter Count} & \textbf{Kernel Size} & \textbf{Strides} & \textbf{Padding} & \textbf{Activation} \\
\midrule
InputLayer & (224, 224, 1) & 0 & - & - & - & - \\
Conv2D & (112, 112, 128) & 2176 & (4, 4) & (2, 2) & same & Linear \\
Activation & (112, 112, 128) & 0 & - & - & - & LeakyReLU \\
Conv2D & (56, 56, 256) & 524288 & (4, 4) & (2, 2) & same & Linear \\
InstanceNormalization & (56, 56, 256) & 512 & - & - & - & - \\
Activation & (56, 56, 256) & 0 & - & - & - & LeakyReLU \\
Conv2D & (28, 28, 512) & 2097152 & (4, 4) & (2, 2) & same & Linear \\
InstanceNormalization & (28, 28, 512) & 1024 & - & - & - & - \\
Activation & (28, 28, 512) & 0 & - & - & - & LeakyReLU \\
Conv2D & (28, 28, 1024) & 8388608 & (4, 4) & (1, 1) & same & Linear \\
InstanceNormalization & (28, 28, 1024) & 2048 & - & - & - & - \\
Activation & (28, 28, 1024) & 0 & - & - & - & LeakyReLU \\
Conv2D & (28, 28, 1) & 16385 & (4, 4) & (1, 1) & same & Linear \\
\bottomrule
\end{tabularx}
\end{table}

\begin{table}[h!t]
\caption{Architecture blocks (Residual, Downsampling, Upsampling) used to construct both generator and discriminator architectures.}
\label{table:blocksarch}
\scriptsize
\begin{tabularx}{\textwidth}{lXXXX}
\toprule
\textbf{Block Type} & \textbf{Layer (type)} & \textbf{Padding} & \textbf{Kernel / Stride} & \textbf{Activation} \\ 
\midrule
Residual & Conv2D & Same & (3,3) / (1,1) & - \\
& InstanceNormalization & - & - & - \\
& ReLU & - & - & ReLU \\
& Conv2D & Same & (3,3) / (1,1) & - \\
& InstanceNormalization & - & - & - \\
& Add (Skip Connection) & - & - & - \\
\midrule
Downsampling & Conv2D & Same & (3,3) / (2,2) & - \\
& InstanceNormalization & - & - & - \\
& ReLU & - & - & ReLU \\
\midrule
Upsampling & Conv2DTranspose & Same & (3,3) / (2,2) & - \\
& InstanceNormalization & - & - & - \\
& ReLU & - & - & ReLU \\
\bottomrule
\end{tabularx}
\end{table}





\end{document}